\documentclass[preprint,aps,superscriptaddress]{revtex4}
\usepackage{graphicx}
\usepackage{dcolumn}
\usepackage{bm}
\usepackage{times}

\begin{document}

\title{Experimental demonstration of photonic crystal waveplates}
\author{D. R. Solli}
\affiliation{Department of Physics, University of California, Berkeley, CA 94720-7300.}
\author{C. F. McCormick}
\affiliation{Department of Physics, University of California, Berkeley, CA 94720-7300.}
\author{R. Y. Chiao}
\affiliation{Department of Physics, University of California, Berkeley, CA 94720-7300.}
\author{J. M. Hickmann}
\affiliation{Department of Physics, University of California, Berkeley, CA 94720-7300.}
\affiliation{Departamento de F\'{\i}sica, Universidade Federal de Alagoas, Cidade
Universit\'{a}ria, 57072-970, Macei\'{o}, AL, Brazil.}

\begin{abstract}
We have constructed and experimentally tested a microwave half waveplate
using the dispersive birefringent properties of a bulk two-dimensional
photonic crystal away from its band gap. Our waveplate device exhibited a
200:1 polarization contrast, limited by our experimental resolution. We
anticipate that photonic crystal waveplates will have important practical
applications in several areas, including integrated photonic circuits.
\end{abstract}

\maketitle

Photonic crystals are passive, usually dielectric, periodic structures
capable of producing distinctive reflected and transmitted fields from
impinging electromagnetic (EM) waves \cite{Yablonovitch1989}. These
structures are known to selectively forbid the propagation of light within
certain frequency and wavevector bands through interference between multiple
Bragg reflections \cite{Chow2000,Hickmann2002}. This phenomenon has been
described as a photonic band gap in analogy with the electronic band
structures of solid state materials. Although this analogy is relevant, the
scalar approximation used for electronic theory neglects polarization
effects, associated with the vector fields of electromagnetic waves,
essential for photonic crystals. Here we explore some polarization-dependent
effects of photonic crystal structures.

Photonic band gap structures can, in principle, be constructed in
essentially any geometry with periodicity in one, two, or all three
dimensions. Unfortunately, fabrication difficulties increase substantially
as the dimensionality of the periodicity is increased. Among the wide
variety of possible applications based on the band structure of photonic
crystals are novel waveguides \cite{Cregan1999} and the suppression of
spontaneous emission, leading to the possibility of thresholdless lasers 
\cite{Yablonovitch1987}. In addition to the well-known intensity effects, it
has also been shown that photonic crystals have a nontrivial effect on the
phase of EM waves in the band gap \cite{Hache2002,Solli2002}.

Although the band gap characteristics of photonic crystals have received
significant attention, their transmission and especially phase effects in
transparent spectral regions have been much less studied. Furthermore, there
has been relatively little attention paid to birefringent polarization
effects associated with photonic crystals. Most of the existing experimental
and theoretical work in this area has been confined to studies of
polarization behavior in photonic band gap waveguides \cite%
{Ortigosa-Blanch2000,Hansen2001,Kerbage2002} or in photonic crystal lasing
cavities \cite{Noda2001}. There has also been some theoretical and
experimental investigation of polarization and birefringence in nonlinear
fiber Bragg gratings \cite{Slusher2000,Pereira2002}. To date, the
polarization properties of two-dimensional photonic crystals have been
discussed only once \cite{Li2001}, in a paper theoretically predicting the
birefringent behavior characteristic of a waveplate.

It is possible to understand the existence of birefringence in a bulk 2D
photonic crystal based on the anisotropy of the structure. A bulk 2D crystal
is one in which the extruded (nonperiodic) dimension is much longer than the
lattice constant. Since the dielectric interfaces are curved, the Fresnel
coefficients for the two possible polarizations are different at normal
incidence, being cumulative for multiple interfaces. An alternative
explanation is found by considering the transmission characteristics of the
structure. In particular, it has already been shown that the amplitude
transmission of an EM wave through a bulk 2D photonic crystal is
polarization-dependent in the band gap region; the center frequency, depth,
and spectral width of the band gap depend on the polarization \cite%
{Hickmann2002}. Furthermore, it is well known that the absorption lines of a
dissipative EM system can have effects on the phases of transmitted waves,
even for frequencies away from the absorptive spectral regions \cite%
{Jackson1999}. Although dielectric photonic crystals are dissipationless,
the band gaps should create similar dispersive effects, even in transparent
spectral regions \cite{Ditchburn1976,Kittel1996}. Since bulk 2D photonic
crystals have well-defined, but characteristically different band gap
properties for waves of different polarizations, they should also possess
birefringent phase delay properties (i.e., indices of refraction) in
transparent spectral regions. In fact, significant phase birefringence has
already been reported within the band gap of a bulk 2D photonic crystal \cite%
{Solli2002}.

Since Maxwell's equations are scale-invariant, all results presented here
apply across the electromagnetic spectrum. Photonic crystal birefringence
should enable the construction of waveplates in all regions of the EM
spectrum. Particularly in the optical region, this large birefringence could
lead to much more compact waveplates than are currently available. It may
also have important applications in photonic crystal circuits, integrated
nonlinear optical systems based on selectively-defective periodic dielectric
structures which have been proposed as all-optical switches \cite%
{Mingaleev2002}. One could imagine highly integrable and compact
photonic-crystal circuit elements such as polarization discriminators,
polarizing beamsplitters and even optical diodes, based on photonic crystal
waveplates.

In this work, we report the experimental demonstration of a bulk
two-dimensional (2D) photonic crystal waveplate. We have observed a
relatively simple photonic band gap structure controllably rotate the angle
of the plane of polarization of an incident linearly polarized EM wave. The
photonic crystals used in this experiment were fabricated using a method
that we have previously described \cite{Hickmann2002}. In summary, we stack
hollow acrylic pipes in a hexagonal array, and cement them together with an
acrylic solvent glue. The pipes used to construct the crystals used in this
experiment have an outer diameter of 1/2 inch, and the final structures have
air-filling fractions of 0.60. We tested crystals ranging from two to twenty
layers of acrylic pipes. The microwave wavelength scale of these systems
means that we can fabricate high-quality crystals relatively easily.

\begin{figure}
\includegraphics{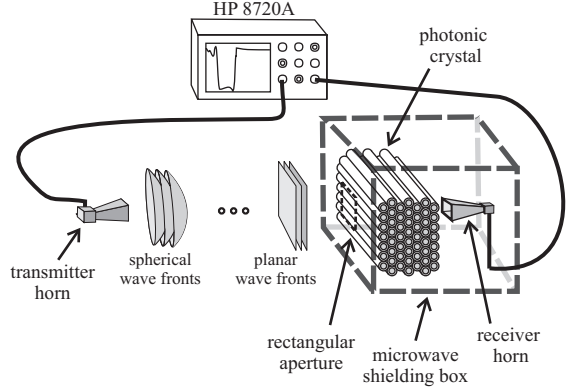}
\caption{Experimental setup for measuring transmission and phase delay of
microwaves passing through the photonic crystal.}
\label{setup}
\end{figure}

Our transmission measurement procedure has also been described elsewhere 
\cite{Hickmann2002}. Microwaves are broadcast from an HP 8720A vector
network analyzer (VNA) and coupled into free space with a
polarization-sensitive antenna (horn). The crystal and receiver horn (also
polarization-sensitive) are placed inside a microwave-shielded box 1.6 m
away with an aperture 14 cm $\times $ 17 cm. These distances are chosen so
that the microwaves arriving at the crystal are effectively plane waves with
a well-defined linear polarization. The crystal is oriented so that the
microwaves are incident in the $\Gamma M$ direction \cite{Foteinopoulou2001}%
. A schematic of the experimental setup is presented in Fig. \ref{setup}.

We have observed a clean exponential formation of the band gap in this
crystal with increasing number of layers \cite{Hickmann2002}. The band gap
center frequency is roughly 11 GHz for transverse magnetic (TM) waves and
10.5 GHz for transverse electric (TE) waves. The band gap depth and width
also depend on polarization. We label the polarizations TE (TM) for the case
in which the electric field is perpendicular (parallel) to the plane of
symmetry of the crystal.

\begin{figure}
\includegraphics{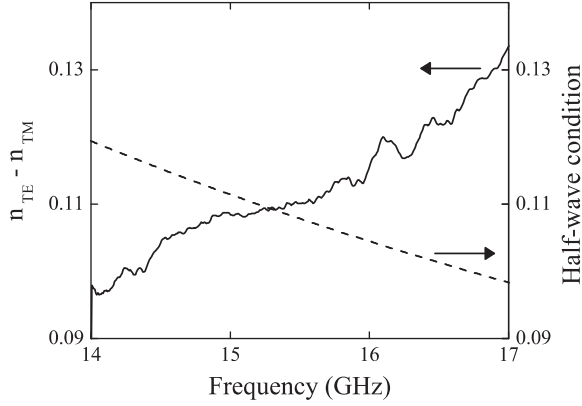}
\caption{(Left axis - solid line) Difference in index of refraction between
TE and TM polarizations ($\Delta n(\protect\omega )$) as a function of
microwave frequency. (Right axis - dashed line) Half-waveplate birefringence
condition, $\frac{\protect\pi c}{\protect\omega L}$.}
\label{index}
\end{figure}

The vector capability of the network analyzer allows us to measure both
transmission and phase delay of the microwaves at the receiver horn. We
initially measured the phase delay in the frequency region of 14 to 20 GHz.
With a simple model of the relation between phase delay and the index of
refraction \cite{Solli2002}, we calculated the index of refraction $n(\omega
)$ for both polarizations of incident microwaves, where\ $\omega $\ is the
angular frequency. For a large range of frequencies (including frequencies
far from the band gap), the two polarizations experience significantly
different values of $n(\omega )$ (see Fig. \ref{index}), demonstrating
birefringence and suggesting that this photonic crystal will behave as a
waveplate for a particular frequency range.

For a given value of $\Delta n(\omega )\equiv n_{TE}(\omega )-n_{TM}(\omega
) $, a zero-order half waveplate shifts the relative phase of the TE and TM
polarizations by $\pi $, leading to the relation

\begin{equation}
|\Delta n(\omega )|=\frac{\pi c}{\omega L}\text{,}  \label{halfwaveplateeqn}
\end{equation}

\noindent where L is the crystal length. Eq. \ref{halfwaveplateeqn} and the
data from Fig. \ref{index} indicate that our eight-layer crystal will act as
a half waveplate for microwaves at 15.3 GHz. To demonstrate this, we first
confirmed that our horns transmitted and received a single well-defined
linear polarization of the electric field; for crossed transmitter and
receiver horns, the power transmission was suppressed by $\geq $ 35 dB
relative to the aligned configuration.

\begin{figure}
\includegraphics{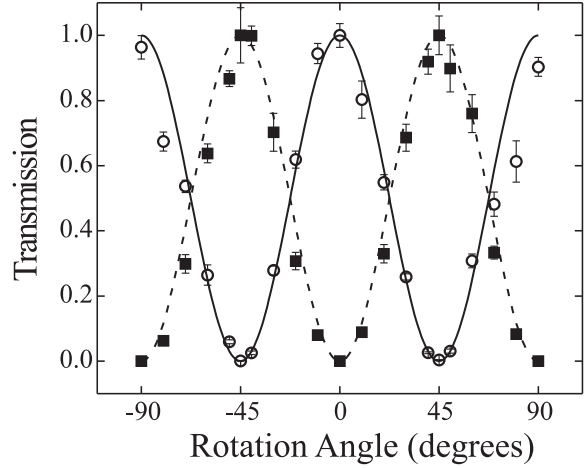}
\caption{Transmission of microwaves averaged over 15.2-15.4 GHz through an eight-layer
hexagonal lattice crystal as a function of the crystal rotation angle. Open
circles: aligned transmitting and receiving horns. Solid squares: crossed
transmitting and receiving horns. Also plotted are the functions $cos^{2}(2%
\protect\theta )$\ (solid line) and $sin^{2}(2\protect\theta )$ (dashed
line) .}
\label{waveplate}
\end{figure}

Next we recorded the power transmission through the crystal averaged over
15.2-15.4 GHz, as it was rotated through $\pm 90^{\circ }$ about the axis of
optical propagation (Fig. \ref{waveplate}). For both crossed (solid squares)
and aligned (open circles) transmitting and receiving horns, the results
show a clear squared-sinusoid (Malus) law for the transmitted power, with
the correct period for a half waveplate. In Fig. \ref{waveplate}, we have
also plotted the functions $cos^{2}(2\theta )$ (solid line) and $%
sin^{2}(2\theta )$ (dashed line), which fit the experimental data extremely
well without the adjustment of any parameter. Since Malus' law holds for
both horn configurations, this effect is clearly due to polarization
rotation and not to an overall drop in transmission through the crystal. The
polarization contrast is better than 200:1 (limited by the experimental
resolution of our setup), indicating that the linear polarization is not
degraded by the crystal.

In conclusion, we have presented the first experimental demonstration of a
photonic crystal waveplate, able to rotate linear polarizations through
arbitrary angles. Any photonic crystal whose band gap properties depend on
polarization should exhibit some type of birefringence. Furthermore, it
should be possible to engineer or tailor the birefringence by tuning to
different transparent frequency bands and/or modifying the relative
properties of the band gaps by changing the filling fraction, lattice
spacing, dielectric composition, or structure geometry. We anticipate that
this type of birefringent behavior in photonic crystals will have
significant applicability, particularly at optical wavelengths in integrated
photonic circuits.

This work was supported by ARO grant number DAAD19-02-1-0276. We thank the
UC Berkeley Astronomy Department, in particular Dr. R. Plambeck, for lending
us the VNA. JMH thanks the support from Instituto do Mil\^{e}nio de Informa%
\c{c}\~{a}o Qu\^{a}ntica, CAPES, CNPq, FAPEAL, PRONEX-NEON, ANP-CTPETRO.

\end{document}